\documentclass{article}
\usepackage{arxiv}
\usepackage{natbib}
\usepackage{graphicx,url}
\usepackage{xurl}
\usepackage{amsmath, listings}
\usepackage[utf8]{inputenc}
\usepackage[english]{babel}
\usepackage{minted}
\usepackage{hyperref}

\title{Implementing the L4S Architecture in the ns-3 Simulator}
\hypersetup{
  colorlinks,
  citecolor=black,
  filecolor=black,
  linkcolor=black,
  urlcolor=black,
  pdflang={en}
}

\author{%
    \\
    \textbf{Maria Eduarda Veras} \\
    Centro de Informática (CIn) \\
    Grupo de Pesquisa em Redes e Telecomunicações\\(GPRT)\\
    Universidade Federal de Pernambuco (UFPE) \\
    Recife, Brasil \\
    \texttt{eduarda.martins@gprt.ufpe.br}
    \and
    \\
    \textbf{Eduardo Freitas} \\
    Centro de Informática (CIn) \\
    Grupo de Pesquisa em Redes e Telecomunicações\\(GPRT)\\
    Universidade Federal de Pernambuco (UFPE) \\
    Recife, Brasil \\
    \texttt{eduardo.freitas@gprt.ufpe.br}
    \and
    \\
    \textbf{Assis T. de Oliveira Filho} \\
    Centro de Informática (CIn) \\
    Grupo de Pesquisa em Redes e Telecomunicações\\(GPRT)\\
    Universidade Federal de Pernambuco (UFPE) \\
    Recife, Brasil \\
    \texttt{assis.tiago@gprt.ufpe.br}
    \and
    \\
    \textbf{Djamel Sadok} \\
    Centro de Informática (CIn) \\
    Grupo de Pesquisa em Redes e Telecomunicações\\(GPRT)\\
    Universidade Federal de Pernambuco (UFPE) \\
    Recife, Brasil \\
    \texttt{jamel@gprt.ufpe.br}
    \and
    \\
    \textbf{Judith Kelner} \\
    Centro de Informática (CIn) \\
    Grupo de Pesquisa em Redes e Telecomunicações\\(GPRT)\\
    Universidade Federal de Pernambuco (UFPE) \\
    Recife, Brasil \\
    \texttt{jk@gprt.ufpe.br}
}
\renewcommand{\shorttitle}{}

\hypersetup{
pdftitle={Implementing the L4S Architecture in the ns-3 Simulator},
pdfsubject={cs.NI},
pdfauthor={Maria Eduarda Veras}
}

\begin{document}

\maketitle

\begin{abstract}
The demand for ultra-low latency in modern applications, such as cloud gaming and augmented reality, has exposed the limitations of traditional congestion control algorithms regarding bufferbloat. The Low Latency, Low Loss, and Scalable Throughput (L4S) architecture addresses this challenge by combining scalable congestion controls, such as TCP Prague, low-latency queue management with prioritization, and Accurate ECN (AccECN) feedback. Although Linux kernel implementations exist, the research community lacks a complete, high-fidelity model within the Network Simulator 3 (ns-3) for reproducible experiments. This paper presents an implementation of end-host protocols for the L4S architecture in ns-3, focusing on the porting of TCP Prague from the Linux kernel (v6.12) and the integration of AccECN signaling. Significant engineering challenges regarding the adaptation of kernel logic are detailed, particularly the reconciliation of Linux’s packet-based arithmetic with ns-3’s byte-based architecture for window management and pacing. Simulation results demonstrate that the proposed model faithfully reproduces the congestion response behaviors observed in real-world testbed scenarios, validating the platform's accuracy. Consequently, this work provides the community with a validated toolset for complex L4S performance evaluations in controlled environments.
\end{abstract}

\section{Introduction}

Over the last decade, a transition has been observed in the profile of network applications, which have moved beyond demanding only high bandwidth to also requiring ultra-low latency. Interactive applications such as cloud gaming, augmented reality, videoconferencing, and control systems for drones and autonomous vehicles require minimal response times to ensure user Quality of Experience (QoE) \cite{rfc9330}. Recent reports from Cisco demonstrate the significant growth of these applications \cite{Cisco2020cisco-annual-internet-report-2018-2023}, highlighting that current network infrastructure needs critical updates to efficiently support these new technologies.

However, understanding the origin of this network latency is crucial. Predominant congestion control algorithms, such as TCP Cubic, standard on Linux, Windows, and Apple network stacks, still rely on the loss detection paradigm \cite{rfc9438}, which remains fundamentally reactive, contrasting with modern proactive requirements. These algorithms tend to occupy network buffers fully until packet loss occurs, a behavior that induces the phenomenon known as bufferbloat: in oversized buffers, packets suffer excessive queueing delays waiting for transmission \cite{5755608}.

To mitigate bufferbloat, the adoption of Active Queue Management (AQM) mechanisms is recommended to manage latency by proactively dropping packets before queue saturation, as detailed in RFC 7567 \cite{rfc7567}. Concurrently, the use of Explicit Congestion Notification (ECN) allows these AQMs to signal congestion in the IP header. Upon detecting this marking, the sender reduces its transmission rate before actual packet loss occurs \cite{rfc3168}. However, a critical limitation exists: TCP's congestion control interprets the standard ECN marking as a loss event, aggressively reducing the congestion window. As a result, throughput is compromised if the AQM marks a large number of packets to maintain low latency, while bufferbloat persists if it marks a small number. Thus, a dilemma is created where it is not possible to obtain low latency and high throughput simultaneously with the current architecture.

Aiming to overcome these limitations, the IETF standardized the Low Latency, Low Loss, and Scalable Throughput (L4S) architecture. The fundamental proposal of L4S is creating an AQM scheme that provides ultra-low queue delay for low-latency traffic, allowing them to receive prioritization. To achieve this, the architecture relies on three pillars: (i) a scalable transport protocol, such as TCP Prague, capable of reacting smoothly to these markings without compromising throughput; (ii) a new feedback scheme, such as Accurate ECN, that enables fine-grain congestion signaling, allowing the network to send frequent and precise feedback to the sender without causing low throughput \cite{rfc9330} \cite{ietf-tcpm-accurate-ecn-34}; and (iii) a coupled queue mechanism which provides low queue delay for the low-latency traffic, while isolating them from the classic traffic, ensuring fair coexistence between both \cite{rfc9332}.

Although the L4S architecture already has mature implementations in the Linux kernel \cite{briscoe2019implementation}, computational simulation remains a fundamental pillar in networking research. As highlighted by Fujimoto et al. \cite{fujimoto2007network}, the increasing complexity of modern protocols and hardware speed make purely analytical techniques insufficient for validating realistic scenarios. Simulation allows overcoming these barriers, ensuring variable control and experiment reproducibility. In this context, the ns-3 network simulator is widely adopted in academia due to its robustness, open-source nature, and, crucially, the high fidelity of its architecture with real network stacks \cite{Riley2010}.

Given this scenario, this work aims to enable the L4S architecture in the ns-3 simulator. To this end, support for Accurate ECN and TCP Prague congestion control was implemented, integrating them with an experimental model of the DualQ Coupled Square AQM mechanism to validate the performance of the complete architecture. It is worth highlighting, to ensure reproducibility and foster further research, the entire implementation is made publicly available as open-source software, providing the community with a verified toolset for future L4S evaluations.


\section{Theoretical Background}
\subsection{L4S Architecture}
To enable traffic coexistence, the L4S architecture needs to separate the low latency traffic from the classic one. First, it uses the IP header ECN field as the primary identifier for each case. While the original ECN standard defined the ECT(0) and ECT(1) codepoints as equivalent and interchangeable \cite{rfc3168}, L4S redefined this semantics to separate flows. In the new architecture proposal, ECT(1) exclusively identifies low-latency traffic (L4S), leaving ECT(0) and Not-ECT to identify classic traffic \cite{rfc9331}. This distinction allows for the correct direction of packets to their respective queues.

\begin{figure}[htb]
    \centering
    \includegraphics[width=0.7\textwidth]{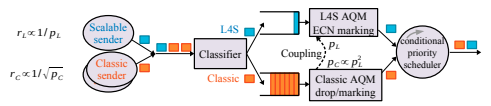}
    \caption{Structure of the DualQ Coupled AQM mechanism. \cite{briscoe2019implementation}}
    \label{fig:aqm}
\end{figure}

To manage these distinct flows, the DualQ Coupled AQM mechanism is employed, as defined in RFC 9332 \cite{rfc9332} and illustrated in Figure \ref{fig:aqm}. Structurally, the mechanism maintains two segregated physical queues: one for classic traffic and another for low-latency traffic (L4S). Since the classic queue does not demand ultra-low latency, it is dimensioned to accommodate data bursts, tolerating longer wait times in favor of throughput. In contrast, the L4S queue is optimized for immediate forwarding. The efficiency of this model is fundamentally based on the coupling between the queues. This mechanism is necessary to ensure fairness; without coupling, L4S traffic could monopolize the bandwidth, leading to the starvation of Classic traffic. Thus, coupling ensures that both flows share the link fairly, regardless of their individual latencies.

However, the feasibility of this low latency depends on a new class of transport algorithms termed ``Scalable''. RFC 9330 \cite{rfc9330} formalizes this definition as, on a steady-state flow, a congestion control where the average time between receiving two congestion signals does not change as flow rate scales, given all other factors remain equal. This means that the recovery time for a scalable congestion control algorithm (CCA) is roughly the same regardless of what rate it is. In contrast, a CCA is unscalable, or classic, when the recovery time increases along with the rate, so a high sending rate will demand a long recovery time. The result is that, for a classic CCA to acquire high bandwidth utilization, it will necessarily spend more time without congestion feedback from the network. In contrast, a scalable CCA can scale its rate without losing feedback on the network congestion.

This scale invariance property ensures that the frequency of ECN markings remains stable regardless of link capacity. This allows AQM to keep queue sizes consistently small without sacrificing throughput, allowing the performance required by the architecture.


\subsection{Accurate ECN}
Accurate ECN (AccECN) is a TCP extension designed to provide fine-grain congestion feedback. Classic ECN (RFC 3168) \cite{rfc3168} is limited to binary feedback, supporting one signal per RTT. This is because if a classic ECN receiver receives a marked packet from the network, it will use the ECE flag on TCP segments to notify the sender of congestion, and it will keep the ECE flag on until the sender tells the receiver it has reacted to the congestion notification. If, during the time the sender took to react to the congestion and tell the receiver, the network kept marking more packets because congestion was high, for example, the sender was unaware of these new marks. Therefore, it is unaware of the \textit{extent} of the congestion, only its existence.

AccECN communicates the exact count of marked packets. This allows congestion control algorithms to obtain a precise view of the network state. The specification \cite{ietf-tcpm-accurate-ecn-34} defines the protocol using either TCP Options or header flags. This work focuses on the essential flag-based implementation, which reuses the AE, CWR, and ECE flags to form the unified ACE field.

Negotiation occurs during the three-way handshake. The client sets the AE, CWR, and ECE flags in the SYN packet. This combination ensures backward compatibility: legacy servers that do not support AccECN will simply ignore the AE flag and interpret the CWR and ECE flags as a standard ECN attempt (or no ECN), triggering a transparent fallback to Classic ECN. If the server supports AccECN, it responds in the SYN/ACK by encoding the IP-ECN field state observed in the received SYN into the TCP header flags, as detailed in Table \ref{tab:accecn_synack}. This serves as an initial integrity check for the path.

\begin{table}[ht]
\centering
\caption{AccECN Feedback Encoding in SYN/ACK}
\label{tab:accecn_synack}
\begin{tabular}{|l|c|c|c|}
\hline
\textbf{IP Marking of Received SYN} & \textbf{AE} & \textbf{CWR} & \textbf{ECE} \\ \hline
Not-ECT (00) & 0 & 1 & 0 \\ \hline
ECT(1) (01)  & 0 & 1 & 1 \\ \hline
ECT(0) (10)  & 1 & 0 & 0 \\ \hline
CE (11)      & 1 & 1 & 0 \\ \hline
\end{tabular}
\end{table}

After the connection is established, the ACE field operates as a 3-bit circular counter. Since this field cannot transmit the absolute total of marked packets, the endpoints maintain state variables: the receiver tracks the total marked packets (\texttt{CepR}), and the sender maintains a local estimate (\texttt{CepS}). For every packet received with the CE flag, the receiver increments \texttt{CepR} and writes the value $\texttt{CepR} \pmod 8$ into the ACE field of the ACK. The sender then decodes this feedback to determine the number of new marks since the last ACK. This is done by calculating the difference (\texttt{delta}) between the received ACE value and the local \texttt{CepS}:

\begin{equation}
    \text{delta} = (ACE_{field} + 8 - (\texttt{CepS} \pmod 8)) \pmod 8
    \label{eq:acc_ecn_delta}
\end{equation}

To handle cases where ACK loss causes the counter to wrap multiple times, the algorithm compares \texttt{delta} against $N$ (the number of acknowledged bytes/segments). If $N \ge 8$, creating ambiguity, the algorithm adopts a conservative approach assuming full congestion:

\begin{equation}
    \text{delta}_{safer} = N - ((N - \text{delta}) \pmod 8)
    \label{eq:delta_safer}
\end{equation}

Finally, the sender updates its local counter ($\texttt{CepS} \leftarrow \texttt{CepS} + \text{delta}$), synchronizing its view with the receiver.

\subsection{TCP Prague}
As discussed, L4S demands a scalable congestion control to properly react to frequent congestion notifications. The ``official'' CCA for this matter is TCP Prague. TCP Prague adapts the Data Center TCP (DCTCP) algorithm \cite{rfc8257} for the public Internet by implementing the ``Prague Requirements'' \cite{briscoe2019implementation}. While inheriting DCTCP's proportional congestion response mechanism, Prague modifies some aspects to ensure coexistence with classic traffic.

The core mechanism relies on a state variable $\alpha$ (Alpha), an Exponential Weighted Moving Average (EWMA) of the marked byte fraction. Unlike DCTCP, which couples updates to the flow's RTT, Prague updates $\alpha$ based on a fixed reference time interval (Target RTT). This decoupling ensures that flows with different physical latencies process feedback over a consistent time scale. When the interval elapses, $\alpha$ is updated as follows:

\begin{equation}
    \alpha \leftarrow (1 - g) \times \alpha + g \times F \label{eq:alpha_update}
\end{equation}

\noindent where $g$ is the filter gain and $F$ is the fraction of marked bytes in the last interval. Consequently, the congestion window is reduced proportionally rather than halving:

\begin{equation}
    cwnd \leftarrow cwnd \times \left(1 - \frac{\alpha}{2}\right) \label{eq:prague_reduction}
\end{equation}

To support this fine-grained control, TCP Prague implements a Fractional Congestion Window. Standard integer-based TCP implementations introduce quantization errors when handling small reductions (e.g., reducing 0.3 segments). Prague maintains the window in fixed-point arithmetic, allowing sub-segment adjustments and ensuring the mathematical smoothness required by Eq. \ref{eq:prague_reduction}.

Finally, strict adherence to Pacing is mandatory. In L4S architectures, where queues operate with low marking thresholds, the packet bursts typical of ACK-clocked transmissions would cause instant buffer filling and ``false positive'' congestion marks. TCP Prague calculates a target rate and enforces precise inter-packet gaps, ensuring that markings reflect actual link saturation rather than transient sender dynamics.

\section{Related Work}
The primary algorithm reference for the L4S architecture resides in the TCP Prague implementation within the Linux kernel, maintained by the L4S project team \cite{linux_prague}. This version serves as the operational gold standard, integrating full support for AccECN signaling and the necessary modifications to the TCP state machine. The Linux code defines the expected behavior of the algorithm in real-world environments, serving as the primary model for validating compliance with the architecture's RFCs.

In parallel, UDP Prague \cite{udp_prague} was developed as a user-space implementation that applies the dynamics of the Prague algorithm over the UDP protocol. The objective of this approach is to isolate the congestion control logic from the complexities inherent in the conventional TCP stack, such as retransmissions and packet ordering. Although simplified, this implementation is crucial for the educational understanding of the algorithm, allowing for the isolated observation of how AccECN feedback influences the congestion window update and sending rate adjustment, serving as a ``white-box'' tool for studying protocol dynamics.

Regarding network simulation, a previous port of TCP Prague to ns-3 was developed during Google Summer of Code 2020 \cite{gsoc2020_prague}. While this initiative established an initial baseline, the implementation presented critical architectural divergences relative to the Linux reference, resulting in fidelity gaps. A significant limitation was the severe approximation in the treatment of marked packets (assuming 100\% marking within a window upon detecting a single CE bit) and the absence of the fractional window concept, which prevented the smooth throughput increases characteristic of Prague. Such limitations motivated the need for refinements in the simulation model to align it with the expected behavior for L4S.

\section{Implementations}
The implementation of the L4S architecture was developed upon the ns-3 network simulator, version 3.46.1. The modifications focused on the \texttt{src/internet} module, which implements the TCP/IP protocol stack. To provide context for the changes described in the subsequent sections, it is important to understand the responsibility of the main simulator classes extended in this work:

\begin{itemize}
    \item \texttt{TcpHeader}: Class defining the bit structure of the TCP header. It is responsible for the serialization and deserialization of fields (such as ports, flags, and windows) traversing the simulated network.

    \item \texttt{TcpSocketState}: Class acting as a data container, storing dynamic connection variables (such as the congestion window, cwnd, and RTT estimates).

    \item \texttt{TcpSocketBase}: Acts as the protocol's central state machine. This class manages the connection lifecycle (\textit{Handshake, Established, Closed}), ACK processing, and retransmission logic.

    \item  \texttt{TcpCongestionOps}: Abstract interface defining the behavior of congestion control algorithms (such as Reno, Cubic, or BBR). The new algorithm (Prague) was implemented by inheriting from this class.
\end{itemize}

This modular structure allowed for decoupling the signaling logic (Accurate ECN) from the congestion control logic (TCP Prague), ensuring an extensible and organized architecture. This is of major importance because, as in the official Linux code, AccECN is expected to work independently of the congestion control. So, a TCP Prague server should work appropriately with a TCP Cubic client, as long as the client has support for AccECN

\subsection{TCP Header Extension}
Implementing AccECN support in the \texttt{TcpHeader} class required expanding the \textit{flags} storage structure. Originally, the simulator used an 8-bit variable (\texttt{uint8\_t}), whose address space was fully occupied by standard flags (SYN, ACK, FIN, etc.). To enable the inclusion of the AE \textit{flag}, the \texttt{m\_flags} variable was redefined as a 16-bit integer (\texttt{uint16\_t}), allowing the extension of the \texttt{Flags\_t} enumeration to allocate the value 256 to the new signal, as detailed in Code \ref{code:tcp_flags}. 

\begin{listing}[!ht]
\begin{lstlisting}[language=C++]
    enum Flags_t
    {
        NONE = 0,  //!< No flags
        FIN = 1,   //!< FIN
        SYN = 2,   //!< SYN
        RST = 4,   //!< Reset
        PSH = 8,   //!< Push
        ACK = 16,  //!< Ack
        URG = 32,  //!< Urgent
        ECE = 64,  //!< ECE
        CWR = 128, //!< CWR
        AE = 256   //!< AE
    };
\end{lstlisting}
\caption{Redefinition of Flags Enumeration in TcpHeader}
\label{code:tcp_flags}
\end{listing}

This change also required a specific adjustment in the \texttt{tcp-general-test} module. The \texttt{SendEmptyPacket} method, which originally received \textit{flags} as \texttt{uint8\_t}, had to have its signature updated to \texttt{uint16\_t} to maintain compatibility with the new header definition and ensure the simulator compiled correctly.

\subsection{TCP Core Modification}
 Integrating AccECN and L4S support required interventions in the core classes of the ns-3 TCP stack. While the \texttt{TcpHeader} class provides the data structure and \texttt{TcpPrague} the decision intelligence, it is within the protocol core, composed of \texttt{TcpSocketState} and \texttt{TcpSocketBase}, that state management and effective processing of the congestion signal occur.

The \texttt{TcpSocketState} class, responsible for maintaining the Transmission Control Block (TCB), was extended to accommodate the data structures required by the AccECN feedback mechanism. The alteration consisted of adding persistent cumulative counters, essential for the sender and receiver to maintain a synchronized view of congestion events. The main variables introduced were:

\begin{itemize}
\item \texttt{m\_cepS}: Cumulative counter tracking the total number of CE-marked packets sent or echoed. This variable acts as the sender's local reference to validate received feedback.
\item \texttt{m\_cepR}: Variable maintained on the receiver side to account for the total volume of CE marks received from the network, the value of which is encoded in the ACE field of the return TCP header.
\item \texttt{m\_delta}: Transient state variable storing the net difference of CE marks between the current ACK and the immediately preceding one. This value quantifies the instantaneous intensity of congestion and serves as the primary input for the Prague algorithm.
\end{itemize}

In the \texttt{TcpSocketBase} class, which implements the TCP finite state machine, modifications focused on capability negotiation and feedback interpretation were necessary. During the connection establishment phase (\textit{Handshake}), the \texttt{SendSyn} method was updated to include AccECN support negotiation. The logic for sending SYN and SYN-ACK packets was adjusted to set the AE, CWR, and ECE \textit{flags} according to the protocol specification. Only after bilateral confirmation of support, verified through flag inspection in the peer's response, is L4S mode activated on the connection, ensuring interoperability with legacy implementations.

In the data transfer phase (\textit{Established}), the ACK processing routine was expanded to decode the ACE field of the TCP header. Upon each received confirmation, \texttt{TcpSocketBase} compares the remote ACE counter value with its local state (\texttt{m\_cepS}), calculating the congestion increment (\texttt{delta}).

This socket-level processing is fundamental to decouple header reading from control logic. \texttt{TcpSocketBase} abstracts the complexity of AccECN signaling and exposes only the processed value (\texttt{m\_delta}) to the \texttt{TcpCongestionOps} interface, allowing the \texttt{TcpPrague} algorithm to directly consume congestion information without needing to manipulate raw protocol bits.

\subsection{TCP Prague Addition}
The introduction of L4S support in the simulator was consolidated through the \texttt{TcpPrague} class. Following the architectural pattern of modern ns-3 implementations, such as BBR and DCTCP, this class inherits from the \texttt{TcpCongestionOps} interface, ensuring compatibility with the existing TCP stack. The class was designed to expose its internal variables, such as control gain and target RTT, through the attribute system (\texttt{TypeId}), allowing flexible customization of simulation parameters without the need for recompilation.

The implemented control logic is strictly based on the Linux kernel source code, specifically version \texttt{l4steam-6.12-y}. The development process consisted of a systematic mapping of kernel functions to the simulator environment (C++). During this process, it was observed that most control functions, such as those responsible for calculating RTT estimates or state management (\texttt{EnterLoss}, \texttt{EnterCwr}), could be transposed with null or minimal alteration, preserving the exact behavior of the reference.

One of the major technical challenges faced during implementation was the fundamental divergence in congestion window representation. While the Linux kernel manages both the main window (\texttt{cwnd}) and its fractional part (\texttt{frac\_cwnd}) in segment units, ns-3 operates with these variables in bytes. This difference required significant effort to ensure the logical equivalence of mathematical operations. In Linux, control arithmetic extensively uses fixed-point operations for performance optimization. In the simulator, the choice was made to simplify this layer using floating-point types (\texttt{double}), which eliminated bit-shifting complexity while maintaining necessary precision.

However, the most critical adaptation involved rounding logic and sending availability. The original algorithm on Linux adopts an aggressive rounding policy: a window of, for example, 2.01 segments is effectively treated as 3 segments, allowing the immediate transmission of the third packet. Since ns-3 treats the window as a continuum of bytes, direct implementation of this logic was not trivial. The adopted solution consisted of implementing a two-step conversion mechanism within update functions: the current window in bytes is temporarily converted to a segment count representation, the Linux-characteristic upper rounding logic is applied (e.g., $\lceil 2.01 \rceil = 3$), and the result is converted back to bytes.

Another aspect of the adaptation refers to the sensing and state variable update logic, specifically in calculating the Exponential Weighted Moving Average (EWMA) for the congestion fraction (\texttt{alpha}) and RTT smoothing. On Linux, variables like \texttt{hsrtt} are stored as scaled integers (multiplied by 128) to allow fast updates via bit-shifting. In ns-3, where model clarity and temporal precision are prioritized, the integer scaling layer was removed in favor of native types (\texttt{double} and \texttt{Time}). The logic was then translated to the standard algebraic EWMA formula:
\begin{equation}
    New = Old + (Sample - Old)/128
\end{equation}
This modification maintains the original statistical weight of the sample ($\approx0.0078$), ensuring that the digital filter's convergence dynamics are identical to the kernel's, but utilizing the simulator's type infrastructure.

Concurrently with window and sensing management, the implementation of the \textit{Pacing} mechanism proved indispensable for the effectiveness of L4S support, aiming to mitigate micro-bursts that could distort congestion signaling. Porting the \texttt{prague\_update\_pacing\_rate} function demanded a reconciliation between kernel and simulator typings. While Linux handles the sending rate as a scalar value in bytes per second, ns-3 uses the \texttt{DataRate} abstraction, oriented in bits per second, requiring the constant application of dimensional conversion factors.

In addition to the unit divergence, the rate scaling logic needed adjustment to the simulator's data granularity. In the reference algorithm, the target rate is derived directly from the number of packets in flight ($R_{base} \times N_{packets}$). To reproduce this behavior in ns-3, where tracking occurs in bytes, a normalization of the \texttt{m\_bytesInFlight} variable by the segment size (\texttt{m\_segmentSize}) was implemented. This adaptation ensured that the sending rate aggressiveness remained mathematically equivalent to the kernel's, even preserving the rate doubling logic during \textit{Slow Start}, although hardware optimization details, such as TSO burst sizing, were abstracted in this version.

\section{Evaluation}

\begin{figure}[htbp]
\centering
\includegraphics[width=0.55\textwidth]{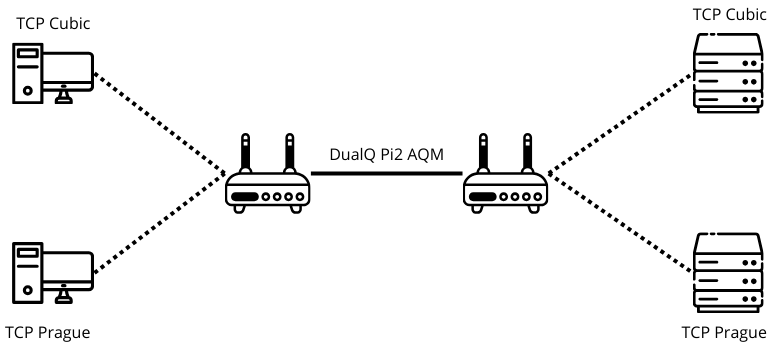}
\caption{Network topology used in the experiments.}
\label{fig:topology}
\end{figure}

To evaluate L4S and TCP Prague in ns-3, the Dumbbell topology in Figure \ref{fig:topology} was used, simulating downlink traffic from right-side servers to left-side clients. Dashed lines represent ideal access links (1 Gbps, 0 ms delay), while the solid central link serves as the controlled bottleneck. This setup purposefully isolates congestion at the central link to analyze queue management performance.

This central bottleneck was subjected to two bandwidth and delay configurations, 100 Mbps + 5 ms, and 10 Mbps + 30 ms, to cover different Bandwidth-Delay Products (BDP). Queue management at this point employs the DualQ Coupled AQM mechanism, specifically, the DualPI2. This queue discipline was installed on the network interface of the router on the right (server side) facing the router on the left, the exact point where downlink packets queue up before traversing the bottleneck.

Since native support for DualPI2 is not yet in the main simulator version, a community-developed experimental implementation tested for IETF compliance was adopted\footnote{DualQ AQM implementation used: \url{https://gitlab.com/tomhenderson/ns-3-dev/-/blob/dual-queue/src/traffic-control/model/dualq-coupled-pi2-queue-disc.cc}}. Although the AQM presents divergences from the Linux DualPI2 version it is considered a valid L4S AQM since it follows the RFC's specifications. To ensure the closest representativeness, the configurable parameters were aligned with the Linux kernel reference implementation (\texttt{sch\_dualpi2}). Furthermore, a minimum queue length check was introduced to comply with RFC 9332 recommendations regarding serialization delay in low-bandwidth links. To saturate the network, the \texttt{BulkSendApplication} was used to generate continuous flows of TCP Prague and TCP Cubic competing for 60 seconds.

Result validation follows a rigorous statistical approach. Thirty independent replications were performed for each scenario. To ensure statistical independence between runs while maintaining experiment reproducibility, the global random number generator seed was fixed, and only the \textit{Run Number} was varied. The reported metrics for throughput, RTT, congestion window (cwnd), queuing delay (sojourn time), coupled probability, and congestion markings represent the sample mean, illustrated with 95\% confidence intervals. To assess coexistence, Raj Jain's Fairness Index is calculated for each replication based on the average throughput of the flows, and the final result represents the mean index across all runs.

To validate our TCP Prague implementation, we replicate the simulated scenario in a real physical testbed. We use the same bottleneck configuration and topology as in ns-3. We use two dedicated hypervisors, each with two virtual machines that represent two clients and two servers. The two hypervisors are connected via two real routers that emulate the bottleneck. All NICs are gigabit Ethernet, and the routers have an Intel Core i5 3.33 GHz 4 cores and 8 GiB of RAM. The operating systems are different simply due to hardware compatibility on the routers, which were more stable and simpler to configure with Debian 13, whereas the VMs use Ubuntu 22.04. Our VMs have one vCPU Intel Xeon E-2434 and 4 GiB of RAM each. We use DualPI2 and TCP Prague version  6.6.114-a76b708b2-l4steam-116 as our L4S software, both from the official L4S repository \cite{linux_prague}.

\section{Results}

\begin{figure}[htbp]
\centering
\includegraphics[width=0.9\textwidth]{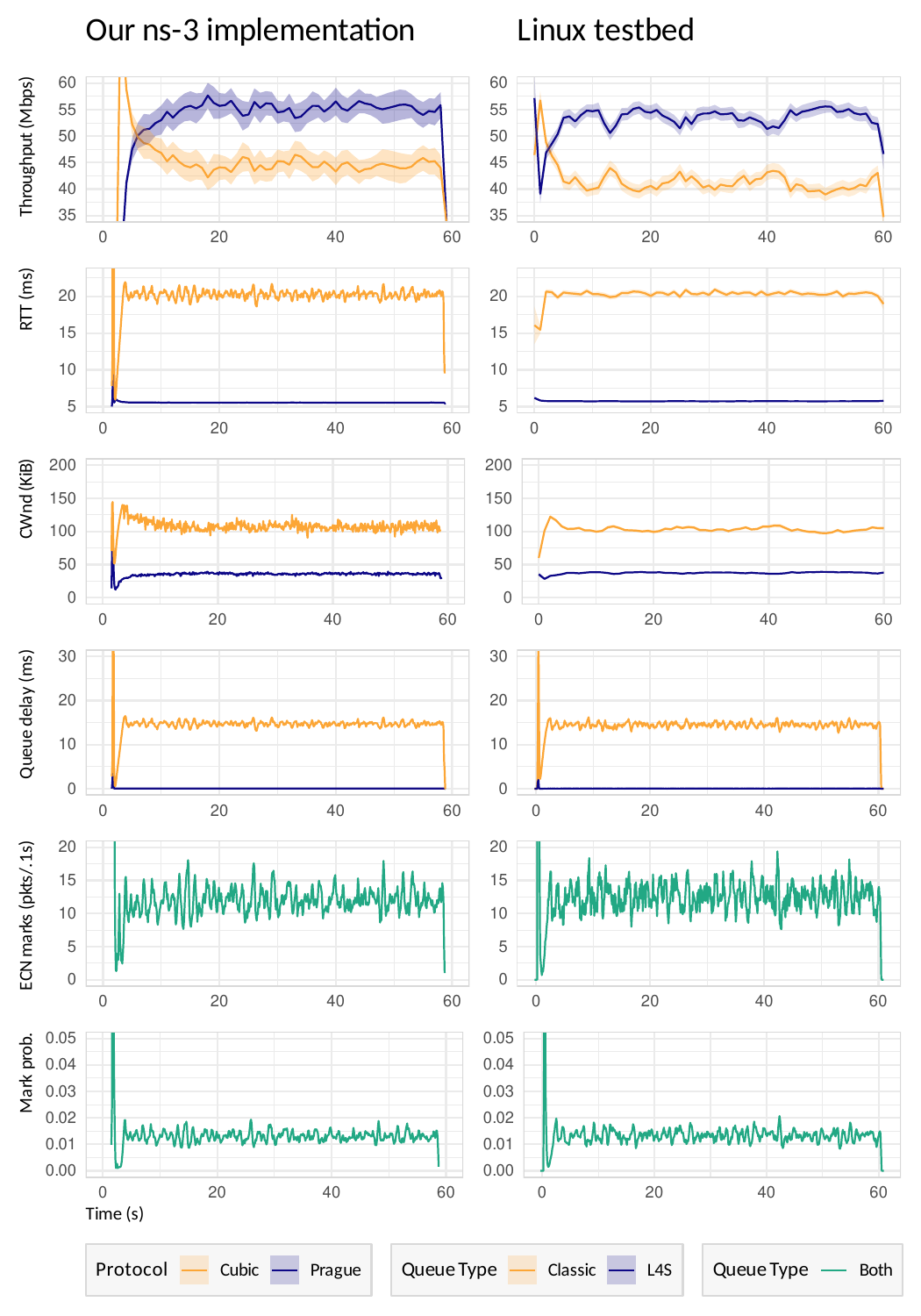}
\caption{Results for the bottleneck of 100 Mbps and 5 ms RTT}
\label{fig:results-scenario1}
\end{figure}

\begin{figure}[htbp]
\centering
\includegraphics[width=0.9\textwidth]{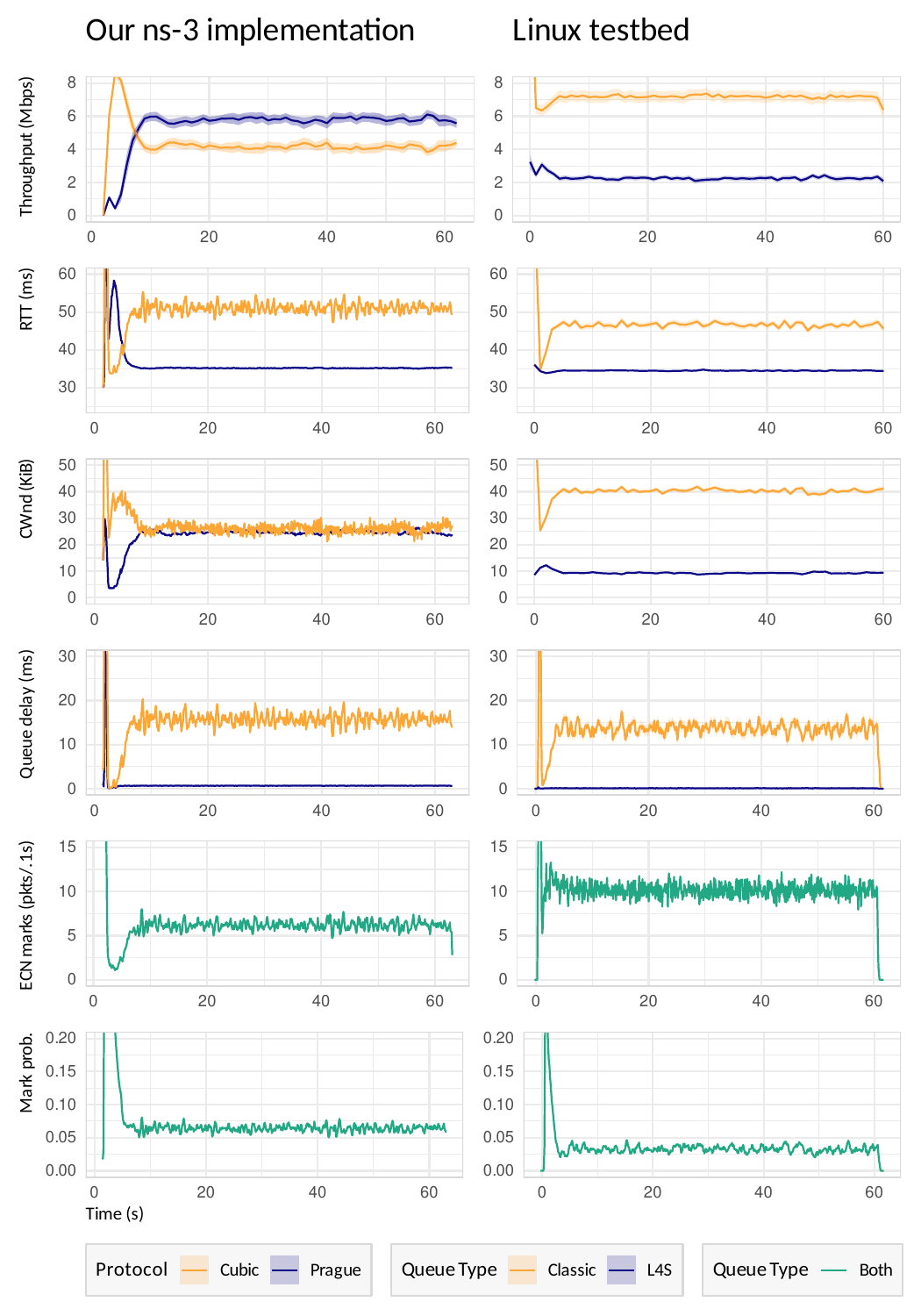}
\caption{Results for the bottleneck of 10 Mbps and 30 ms RTT}
\label{fig:results-scenario2}
\end{figure}

To validate our ns-3 implementation, the simulation results for the 100 Mbps + 5 ms scenario show graphs closely matching the Linux testbed, as illustrated in Figure \ref{fig:results-scenario1}. Both TCP Prague and TCP Cubic exhibit similar behavior patterns in RTT and congestion window dynamics. The results demonstrate RTT stability around 20 ms for Cubic and 6 ms for Prague, evidencing the low latency achieved by Prague. Furthermore, the congestion window of both protocols shows strong agreement, illustrating how Prague maintains low delay without starving the competing flow since the Cubic window remains approximately twice as large.

The queue statistics show strong similarity in the 100 Mbps + 5 ms scenario, with consistent behavior in both ECN marking and marking probability graphs. Given the lower network utilization in this scenario, the AQM successfully maintains coexistence between flows, with Jain's Fairness Index of 0.9927 ($\pm$ 0.0042), in addition to its validated behavior in relation to the Linux test environment. Finally, since throughput is directly derived from the congestion window, the 100 Mbps + 5 ms scenario exhibits nearly identical results across both implementations, consistent with the previously observed metrics. Prague stabilizes around 55 Mbps, while Cubic achieves $\approx$45 Mbps.

For the second scenario, with results shown in Figure \ref{fig:results-scenario2}, we see a slight divergence from the Linux testbed, although the differences still comply with the Prague Requirements. The Prague flow sustained low queue delay and RTT throughout the transmission. It also ensured fair coexistence with Cubic, evidenced by a Jain’s Fairness Index of 0.9876 ($\pm$ 0.0056) and the absence of starvation.

However, the divergence occurs in the congestion window, which can be attributed to the DualPI2 implementation. The AQM adopted in our experiments ``allows'' a higher queue delay on both queues, which can be seen in the graph for the classic queue. The ns-3 version has an overall mean of 15.6 ms, while in the Linux testbed it has 12.7 ms. The same results are reflected in the L4S queue, with an overall mean of 0.84 ms ($\pm$3.33 ms) for ns-3 and 0.09 ms ($\pm$0.28 ms) for the Linux testbed. The overall higher queue delay, however, still does not result in higher packet marking in this scenario, which in turn results in the servers slowly converging their congestion window until the queue stabilizes the packet marking to maintain Prague's fair share of bandwidth utilization. We consider that the main contributor to this issue is the usage of a time-shifted scheduler on the AQM instead of the weighted round-robin approach as in the Linux kernel. Previous literature \cite{deschepper2022dualqueuecoupledaqm} states that such a scheduler can \textit{leak} classic queue delay into the L4S queue, resulting in higher queue delay allowance for both queues. In future research, we aim at providing a DualPI2 AQM algorithm that better reflects the approaches on the Linux kernel for further improved precision.

\section{Conclusion and Future Work}
This work presented an implementation of the L4S architecture in the ns-3 network simulator, including the TCP Prague congestion control algorithm and Accurate ECN. The implementation was validated against a Linux testbed across two network scenarios with different bandwidth-delay product characteristics. Results demonstrate strong agreement between ns-3 and Linux implementations, particularly in RTT behavior and queue delay isolation, confirming that Prague successfully achieves low latency while coexisting with classic Cubic flows.

Our analysis identified a higher queue delay allowance in the AQM implementation, particularly in the lower bandwidth scenario. The reduced ECN marking rate causes Prague to claim a higher bandwidth share than observed in the Linux testbed. Despite this limitation, the priority mechanism correctly isolates L4S traffic from classic queue buildup, maintaining low queue delays across both implementations with fair coexistence, with Jain's Fairness of 0.99 and 0.98 for scenarios 1 and 2, respectively.

The presented ns-3 implementation provides researchers with a validated tool for investigating L4S behavior in controlled simulation environments. Future work will focus on calibrating the DualQ marking parameters and scheduler algorithm and extending validation to scenarios with multiple concurrent flows and varying network topologies.

\section*{Code availability}
\label{sec:source-code}
All source code that we developed is publicly available in our project repository as ns-3 patches at \href{https://github.com/GPRT/l4s-for-ns3}{https://github.com/GPRT/l4s-for-ns3}. We are currently developing the DualPI2 code and improving our tests and experimentation.

\section*{Acknowledgment}
\label{sec:acknowledgment}
This research was supported by \textit{Fundação de Amparo à Pesquisa do Estado de São Paulo (FAPESP)}.

\section*{Declaration of artificial intelligence usage}
\label{sec:aideclaration}
The Google Gemini tool was utilized in this work exclusively to assist with the translation of specific sections and for linguistic refinement. It is explicitly declared that no scientific content, data simulation, coding, or analysis was generated by artificial intelligence.

\bibliographystyle{unsrt}
\bibliography{references}

@unpublished{deschepper2022dualqueuecoupledaqm,
  title =        {Dual Queue Coupled AQM: Deployable Very Low Queuing Delay for All},
  author =       {Koen De Schepper and Olga Albisser and Olivier Tilmans and Bob Briscoe},
  year =         2022,
  eprint =       {2209.01078},
  archivePrefix ={arXiv},
  primaryClass = {cs.NI},
  url =          {https://arxiv.org/abs/2209.01078},
}

@techreport{Cisco2020cisco-annual-internet-report-2018-2023,
  title = "Cisco Annual Internet Report (2018–2023)",
  author = "",
  institution = "Cisco",
  url = "https://www.cisco.com/c/en/us/solutions/collateral/executive-perspectives/annual-internet-report/white-paper-c11-741490.html",
  year = "2020"
}

@misc{rfc3168,
    series =    {Request for Comments},
    number =    3168,
    howpublished =  {RFC 3168},
    publisher = {RFC Editor},
    doi =       {10.17487/RFC3168},
    url =       {https://www.rfc-editor.org/info/rfc3168},
    author =    {Sally Floyd and Dr. K. K. Ramakrishnan and David L. Black},
    title =     {{The Addition of Explicit Congestion Notification (ECN) to IP}},
    pagetotal = 63,
    year =      2001,
    month =     sep,
}

@misc{rfc9330,
    series =    {Request for Comments},
    number =    9330,
    howpublished =  {RFC 9330},
    publisher = {RFC Editor},
    doi =       {10.17487/RFC9330},
    url =       {https://www.rfc-editor.org/info/rfc9330},
    author =    {Bob Briscoe and Koen De Schepper and Marcelo Bagnulo and Greg White},
    title =     {{Low Latency, Low Loss, and Scalable Throughput (L4S) Internet Service: Architecture}},
    pagetotal = 36,
    year =      2023,
    month =     jan,
}

@misc{rfc9331,
    series =    {Request for Comments},
    number =    9331,
    howpublished =  {RFC 9331},
    publisher = {RFC Editor},
    doi =       {10.17487/RFC9331},
    url =       {https://www.rfc-editor.org/info/rfc9331},
    author =    {Koen De Schepper and Bob Briscoe},
    title =     {{The Explicit Congestion Notification (ECN) Protocol for Low Latency, Low Loss, and Scalable Throughput (L4S)}},
    pagetotal = 52,
    year =      2023,
    month =     jan,
}

@techreport{ietf-tcpm-accurate-ecn-34,
    number =    {draft-ietf-tcpm-accurate-ecn-34},
    type =      {Internet-Draft},
    institution =   {Internet Engineering Task Force},
    publisher = {Internet Engineering Task Force},
    note =      {Work in Progress},
    url =       {https://datatracker.ietf.org/doc/draft-ietf-tcpm-accurate-ecn/34/},
    author =    {Bob Briscoe and Mirja Kühlewind and Richard Scheffenegger},
    title =     {{More Accurate Explicit Congestion Notification (AccECN) Feedback in TCP}},
    pagetotal = 73,
    year =      2025,
    month =     mar,
    day =       10,
}

@misc{rfc9332,
    series =    {Request for Comments},
    number =    9332,
    howpublished =  {RFC 9332},
    publisher = {RFC Editor},
    doi =       {10.17487/RFC9332},
    url =       {https://www.rfc-editor.org/info/rfc9332},
    author =    {Koen De Schepper and Bob Briscoe and Greg White},
    title =     {{Dual-Queue Coupled Active Queue Management (AQM) for Low Latency, Low Loss, and Scalable Throughput (L4S)}},
    pagetotal = 52,
    year =      2023,
    month =     jan,
}

@ARTICLE{5755608,
  author={Gettys, Jim},
  journal={IEEE Internet Computing}, 
  title={Bufferbloat: Dark Buffers in the Internet}, 
  year={2011},
  volume={15},
  number={3},
  pages={96-96},
  doi={10.1109/MIC.2011.56}
}

@misc{rfc9438,
    series =    {Request for Comments},
    number =    9438,
    howpublished =  {RFC 9438},
    publisher = {RFC Editor},
    doi =       {10.17487/RFC9438},
    url =       {https://www.rfc-editor.org/info/rfc9438},
    author =    {Lisong Xu and Sangtae Ha and Injong Rhee and Vidhi Goel and Lars Eggert},
    title =     {{CUBIC for Fast and Long-Distance Networks}},
    pagetotal = 28,
    year =      2023,
    month =     aug,
}

@misc{rfc7567,
    series =    {Request for Comments},
    number =    7567,
    howpublished =  {RFC 7567},
    publisher = {RFC Editor},
    doi =       {10.17487/RFC7567},
    url =       {https://www.rfc-editor.org/info/rfc7567},
    author =    {Fred Baker and Gorry Fairhurst},
    title =     {{IETF Recommendations Regarding Active Queue Management}},
    pagetotal = 31,
    year =      2015,
    month =     jul,
}

@Inbook{Riley2010,
author="Riley, George F.
and Henderson, Thomas R.",
title="The ns-3 Network Simulator",
bookTitle="Modeling and Tools for Network Simulation",
year="2010",
publisher="Springer Berlin Heidelberg",
address="Berlin, Heidelberg",
pages="15--34",
isbn="978-3-642-12331-3",
doi="10.1007/978-3-642-12331-3_2",
url="https://doi.org/10.1007/978-3-642-12331-3_2"
}

@book{fujimoto2007network,
  title={Network Simulation},
  author={Fujimoto, R.M. and Perumalla, K.S. and Riley, G.F.},
  isbn={9781598291100},
  series={Synthesis lectures on communication networks},
  url={https://books.google.com.br/books?id=YAFxmw6FHXAC},
  year={2007},
  publisher={Morgan \& Claypool Publishers}
}

@misc{briscoe2019implementation,
  title={Implementing the {TCP} {Prague} {L4S} requirements in {Linux}},
  author={Briscoe, Bob and De Schepper, Koen and Albisser, Olga and Tilmans, Olivier},
  howpublished={Presentation at Netdev 0x13 Conference, Prague},
  year={2019},
  month={March}
}

@misc{rfc8257,
    series =    {Request for Comments},
    number =    8257,
    howpublished =  {RFC 8257},
    publisher = {RFC Editor},
    doi =       {10.17487/RFC8257},
    url =       {https://www.rfc-editor.org/info/rfc8257},
    author =    {Stephen Bensley and Dave Thaler and Praveen Balasubramanian and Lars Eggert and Glenn Judd},
    title =     {{Data Center TCP (DCTCP): TCP Congestion Control for Data Centers}},
    pagetotal = 17,
    year =      2017,
    month =     oct,
}

@misc{linux_prague,
  author = {{L4S Team}},
  title = {TCP Prague Linux Kernel Implementation},
  howpublished = {\url{https://github.com/L4STeam/linux/blob/l4steam-6.12.y/net/ipv4/tcp_prague.c}},
  note = {Accessed: 2026-01-19}
}

@misc{udp_prague,
  author = {{L4S Team}},
  title = {UDP Prague: User-space congestion control implementation},
  howpublished = {\url{https://github.com/L4STeam/udp_prague/}},
  note = {Accessed: 2026-01-19}
}

@misc{gsoc2020_prague,
  author = {Deepak K and Ankit Deepak and Mohit Tahiliani and Vivek Jain and Viyom Mittal},
  title = {TCP Prague implementation for ns-3 (GSoC 2020)},
  howpublished = {\url{https://www.nsnam.org/wiki/GSOC2020Prague}},
  year = {2020},
  note = {Accessed: 2026-01-19}
}

\end{document}